\documentclass[aps,prl,twocolumn,superscriptaddress]{revtex4}

\usepackage{graphicx,color,ulem}

\begin{document}

\title{The crystalline electric field as a probe for long range antiferromagnetic order and superconductivity 
in CeFeAsO$_{1-x}$F$_x$}
\author{Songxue Chi}
\affiliation{Department of Physics and Astronomy, The University of Tennessee, Knoxville,
Tennessee 37996-1200}
\author{D. T. Adroja}
\affiliation{ISIS Facility, Rutherford Appleton Laboratory, Chilton, Didcot, Oxfordshire OX11 0 QX, UK}
\author{T. Guidi}
\affiliation{ISIS Facility, Rutherford Appleton Laboratory, Chilton, Didcot, Oxfordshire OX11 0 QX, UK}
\author{R. Bewley}
\affiliation{ISIS Facility, Rutherford Appleton Laboratory, Chilton, Didcot, Oxfordshire OX11 0 QX, UK}
\author{Shliang Li}
\affiliation{Department of Physics and Astronomy, The University of Tennessee, Knoxville,
Tennessee 37996-1200}
\author{Jun Zhao}
\affiliation{Department of Physics and Astronomy, The University of Tennessee, Knoxville,
Tennessee 37996-1200}
\author{J. W. Lynn}
\affiliation{NIST Center for Neutron Research, National Institute of Standards and
Technology,Gaithersburg, MD 20899-6102}
\author{C. M. Brown}
\affiliation{NIST Center for Neutron Research, National Institute of Standards and
Technology,Gaithersburg, MD 20899-6102}
\author{Y. Qiu}
\affiliation{NIST Center for Neutron Research, National Institute of Standards and
Technology,Gaithersburg, MD 20899-6102}
\affiliation{Department of Materials Science and Engineering, University of Maryland,College, MD 20742}
\author{G. F. Chen}
\affiliation{Institute of
Physics, Chinese Academy of Sciences, P. O. Box 603, Beijing 100080, China }
\author{J. L. Lou}
\affiliation{Institute of
Physics, Chinese Academy of Sciences, P. O. Box 603, Beijing 100080, China }
\author{N. L. Wang}
\affiliation{Institute of
Physics, Chinese Academy of Sciences, P. O. Box 603, Beijing 100080, China }
\author{Pengcheng Dai}
\email{daip@ornl.gov}
\affiliation{Department of Physics and Astronomy, The University of Tennessee, Knoxville,
Tennessee 37996-1200}
\affiliation{Neutron Scattering Science Division, Oak Ridge National Laboratory, Oak
Ridge, Tennessee 37831-6393}

\begin{abstract}
We use inelastic neutron scattering to study the crystalline electric field (CEF)
excitations of Ce$^{3+}$ in CeFeAsO$_{1-x}$F$_{x}$($x=0,0.16$).
For nonsuperconducting CeFeAsO, the Ce CEF levels have three magnetic doublets in the paramagnetic state, but these doublets split into six singlets when Fe ions order antiferromagnetically. For superconducting CeFeAsO$_{0.84}$F$_{0.16}$ ($T_c=41$ K), where the static AF order is suppressed,
the Ce CEF levels have three magnetic doublets at $\hbar\omega=0,18.7,58.4$ meV at all temperatures.  Careful measurements of 
the intrinsic linewidth $\Gamma$ and the   
peak position of the 18.7 meV mode reveal clear anomaly at $T_c$, consistent with a strong enhancement of local magnetic susceptibility $\chi^{\prime\prime}(\hbar\omega)$ below $T_c$. These results suggest that CEF excitations in the rare-earth oxypnictides can be used as a probe of spin dynamics in the nearby FeAs planes.  
\end{abstract}


\maketitle

The rare-earth ($R$) oxypnictides with general formula $R$FeAsO ($R =$ La, Sm, Ce, Nd, and Pr)
are currently attracting much attention due to the discovery of 
high-transition temperature (high-$T_c$) superconductivity in these materials upon chemical
doping \cite{kamihara,chen,gfchen,zaren,hhwen}. Although superconductivity in electron doped LaFeAsO
appears at a moderate superconducting temperature of 28 K \cite{kamihara}, replacing La with other rare earth ions  increases $T_c$ up to 55 K \cite{chen,gfchen,zaren}, making the rare-earth oxypnictides a new class of high-$T_c$ superconductors with critical temperatures only surpassed 
by high-$T_c$ copper oxides.  Since superconductivity in these rare-earth oxypnictides appears after electron-doping to suppress the static antiferromagnetic (AF) order in their parent compounds 
\cite{cruz,zhao,mcguire,chen1,huang}, it is important to determine the influence of magnetic interactions on the superconducting properties. Compared to LaFeAsO, where Fe is the only possible 
ion carrying a significant magnetic moment, the rare-earth oxypnictide with unpaired $4f$ electrons such as  
Ce$^{3+}$ in CeFeAsO offers an unique opportunity to study the interplay between the rare-earth 
and Fe magnetic ions. In particular, by using neutron scattering to study the crystal-electric-field (CEF) excitations of the rare-earth in $R$FeAsO and their doped superconductors, one can determine the electronic ground state of the $R$ ions and therefore understand the low temperature thermodynamic and magnetic properties of these materials \cite{fulde,mesot}. Furthermore, since the $R$ ions are situated near the superconducting FeAs layer, CEF excitations at the $R$ sites are sensitive to the electronic properties of the FeAs layer, and can be used as a probe to study the influence of superconductivity on spin fluctuations in the FeAs plane \cite{osborn,boothroyd,mesot1}.

In this paper, we report inelastic neutron scattering studies of the Ce$^{3+}$ CEF excitations in the AF ordered CeFeAsO and superconducting CeFeAsO$_{0.84}$F$_{0.16}$ ($T_c=41$ K) \cite{zhao}. 
For CeFeAsO, we find that 
the Ce$^{3+}$ CEF levels are composed of 
three magnetic doublets at $\hbar\omega=0, 18.7$, and 60 meV in the paramagnetic state.
When the Fe long range AF order sets in, these doublets are split into six singlets. 
In the case of CeFeAsO$_{0.84}$F$_{0.16}$, although the three doublets are no longer split, 
 the intrinsic linewidth $\Gamma$ and the   
peak position of the $\hbar\omega_1=18.7$ meV mode show a clear anomaly below $T_c$. These results suggest that the linewidth and position of the CEF transitions in the rare-earth oxypnictides are sensitive to the Fe spin ordering, and  
 can be used as a direct probe of the spin dynamics in the nearby FeAs planes.  

Our experiments were carried out using the MERLIN chopper spectrometer at ISIS facility, Rutherford-Appleton Laboratory, Didcot, UK \cite{merlin}; the BT4 Filter Analyzer Neutron Spectrometer (FANS) at the NIST center for neutron research (NCNR) \cite{udovic}; and the NG4 Disk-chopper time-of-flight spectrometer (DCS) at NCNR, Gaithersburg, Maryland. MERLIN is a high count rate, medium energy resolution, direct geometry chopper spectrometer with a large solid angle of position sensitive detectors. FANS is a high count instrument to measure inelastic excitations on powders and DCS is a cold neutron direct geometry chopper spectrometer. Our CeFeAsO and 
CeFeAsO$_{0.84}$F$_{0.16}$ samples were prepared using methods described in Ref. \cite{chen} and their structural/magnetic properties are discussed in Ref. \cite{zhao}. For MERLIN measurements, a low temperature He-4 cryostat and closed cycle He-gas refrigerator were used for the temperature variable. The incident beam energies were $E_i= 7, 10, 30, 70, 105, 300$ meV. To separate the CEF magnetic scattering from phonon excitations, we also measured LaFeAsO and LaFeAsO$_{0.88}$F$_{0.08}$ samples \cite{cruz} as reference materials for phonon subtraction from the Ce samples. The data for CeFeAsO$_{1-x}$F$_{x}$ and LaFeAsO$_{1-x}$F$_{x}$
were converted to absolute units of mb/sr/meV/f.u. by normalizing the scattering from vanadium measurements made with the same incident energies. For the DCS measurements, we used a He-4 cryostat and an incident beam energy of 3.55 meV and for the FANS measurements we used a top-loading He-gas refrigerator.
 
\begin{figure}[tbp]
\includegraphics[scale=.4]{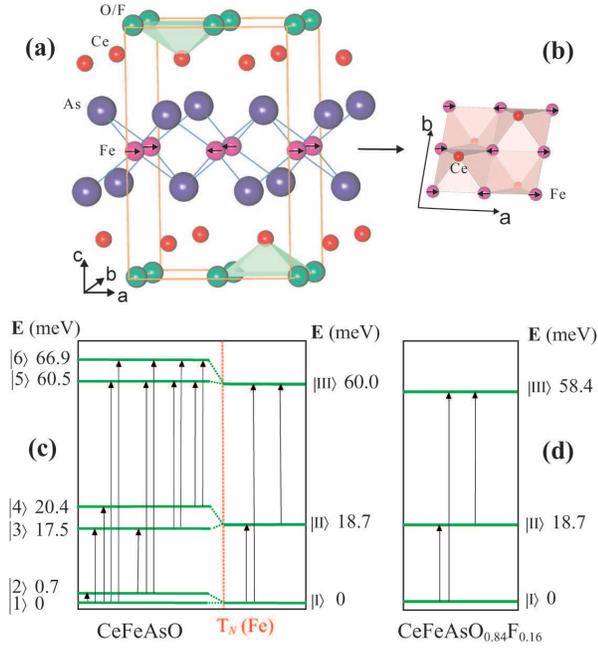}
\caption{(Color online) Summary of the CeFeAsO crystal/magnetic structure and CEF levels determined from our inelastic neutron scattering experiments. (a) The Fe spin ordering in the CeFeAsO chemical
unit cell. (b) The Fe spins in CeFeAsO with respect to the Ce positions. The
Fe moments lie in the $a$-$b$ plane along the $a$-axis and form an
antiferromagnetic collinear spin structure. (c) Ce$^{3+}$ CEF levels in CeFeAsO for temperatures above and below the Fe AF N$\rm \acute{e}$el temperature of $T_N=140$ K \cite{zhao}.  The arrows indicate possible transitions.  (d) Ce$^{3+}$ CEF levels in superconducting CeFeAsO$_{0.84}$F$_{0.16}$ at low temperature. }
\end{figure}

Figure 1(a) shows the position of the Ce ion in the crystal structure environment of CeFeAsO.  Relative to the Fe sublattice, the Ce$^{3+}$ ions are located alternately above and below the (AF ordered) Fe layers as shown in Fig. 1(b). Figures 1(c) and 1(d) summrize the Ce CEF excitation energies determined from our inelastic neutron scattering experiments for CeFeAsO and CeFeAsO$_{0.84}$F$_{0.16}$, respectively. According to neutron powder diffraction experiments \cite{zhao},
both CeFeAsO and CeFeAsO$_{0.84}$F$_{0.16}$ have a tetragonal (space group $P4/nmm$) crystal structure at room temperature.  However, on cooling, CeFeAsO first exhibits a
structural phase transition, changing the crystal symmetry from tetragonal  
to orthorhombic (space group $Cmma$), and then orders antiferromagnetically with a spin structure
as shown in Figs. 1a and 1b \cite{cruz,zhao}; CeFeAsO$_{0.84}$F$_{0.16}$ maintains the tetragonal structure for all temperatures and does not order magnetically above 4 K.  In the tetragonal structure, the Ce atoms are located at the $2c$ crystallographic site which has $C_{4V}$ point symmetry. This gives three non-zero CEF parameters in the crystal field Hamiltonian and its form in Stevens operators formalism is 
$H_{CEF}(C_{4v})=B_2^0O_2^0+B_4^0O_4^0+B_4^4O_4^4$. In the case of the low-temperature orthorhombic structure of CeFeAsO, the Ce atoms are at the $4g$ $(0,1/4,z)$ site which gives local point symmetry of $mm2$ ($C_{2v}$). The resulting Hamiltonian then involves five non-zero crystal field parameters as $H_{CEF}(C_{2v})=B_2^0O_2^0+B_2^2O_2^2+B_4^0O_4^0+B_4^2O_4^2+B_4^4O_4^4$, where $B_n^m$'s are the 
CEF parameters to be determined from the experimental data; and $O_n^m$'s are operator equivalents obtained using the angular momentum operators \cite{fulde}. 

\begin{figure}[tbp]
\includegraphics[scale=.40]{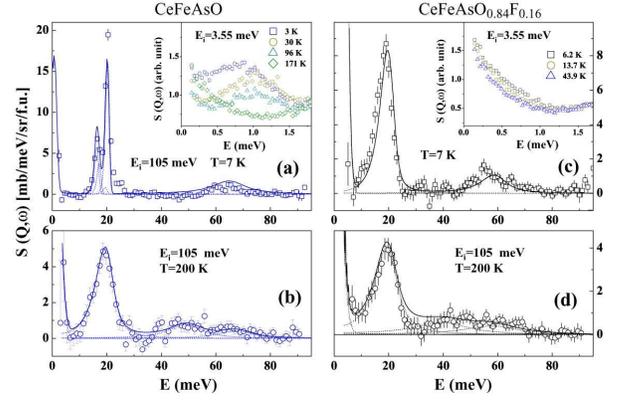}
\caption{(Color online) Temperature dependence of the CEF excitations in 
CeFeAsO and CeFeAsO$_{0.84}$F$_{0.16}$ and our model determination of the CEF levels.
 (a) Ce CEF magnetic excitations in absolute units after subtracting the LaFeAsO phonons. 
 The solid line is our model fit with parameters listed in Table I.  The inset shows
 the raw DCS data integrated from $0<Q<2.2$ \AA$^{-1}$ in arbitrary 
 units as a function of temperature 
 without the LaFeAsO phonon subtraction.
 (b) Ce CEF excitations at 200 K; solid line is our model calculation.
  (c-d) CEF excitations and their temperature dependence for CeFeAsO$_{0.84}$F$_{0.16}$.  The solid
  lines are model calculations.  Inset in (c) shows that the 0.7 meV excitation 
  seen in CeFeAsO is missing in CeFeAsO$_{0.84}$F$_{0.16}$. Error bars represent one standard deviation.}
\end{figure}

\begin{table}[tp]
\caption{Refined $B_n^m$ CEF parameters for CeFeAsO and CeFeAsO$_{0.84}$F$_{0.16}$.}
\label{Table}%
\begin{ruledtabular}
\begin{tabular}{cccccccccc}
        & CeFeAsO$_{0.84}$F$_{0.16}$   & CeFeAsO ($>T_N$) & CeFeAsO ($<T_N$) \\[3pt]
\hline&  \\[3pt]
$B_2^0$ & $2.29\pm0.11$                & $3.19\pm0.15$     & $1.648\pm0.035$  \\[3pt]
$B_4^0$ & $-0.061\pm0.0064$            & $-0.032\pm0.0072$ & $0.1073$         \\[3pt]
$B_2^2$ &                              &                   & $-3.098\pm0.064$  \\[3pt]
$B_4^2$ &                              &                   & $-0.288\pm0.009$  \\[3pt]
$B_4^4$ &  $0.698\pm0.023$             & $0.755\pm0.049$   & $0.591\pm0.018$  \\[3pt]
\end{tabular}
\end{ruledtabular}
\end{table}

We collected neutron scattering data on CeFeAsO and CeFeAsO$_{0.84}$F$_{0.16}$ on MERLIN and DCS spectrometers 
with different incident beam energies to search for Ce$^{3+}$ CEF excitations.
To eliminate phonon scattering, we carried out identical scans using LaFeAsO and LaFeAsO$_{0.88}$F$_{0.08}$ as reference materials.  Figures 2(a) and (b) show the phonon subtracted energy scans 
for CeFeAsO at 7 K and 200 K on MERLIN obtained with $E_i=105$ meV.  The inset of Fig. 2(a) plots similar data taken on DCS with $E_i=3.55$ meV.  Identical scans taken for CeFeAsO$_{0.84}$F$_{0.16}$ are shown in Figs. 2(c) and (d).  We first discuss results on 
the superconducting sample as there are no complications of structural distortion and Fe magnetic order. 
To obtain the $B_n^m$'s CEF parameters in the $H_{CEF}(C_{4v})$ Hamiltonian, we first used the FOCUS program, which has a Monte Carlo search routine, to fit 
the observed two CEF excitations at 18.7 meV and 58.4 meV in Figs. 2(c). We then fit many spectra simualteniously with different incident energies and temperatures using a CEF fit program and the results are plotted as solid lines in Figs. 2(c) and (d) for 7 K and 200 K, respectively. Tables I and II summarize the $B_n^m$'s CEF parameters and wave functions   
for CeFeAsO$_{0.84}$F$_{0.16}$ obtained from those fits.

\begin{figure}[tbp]
\includegraphics[scale=0.45]{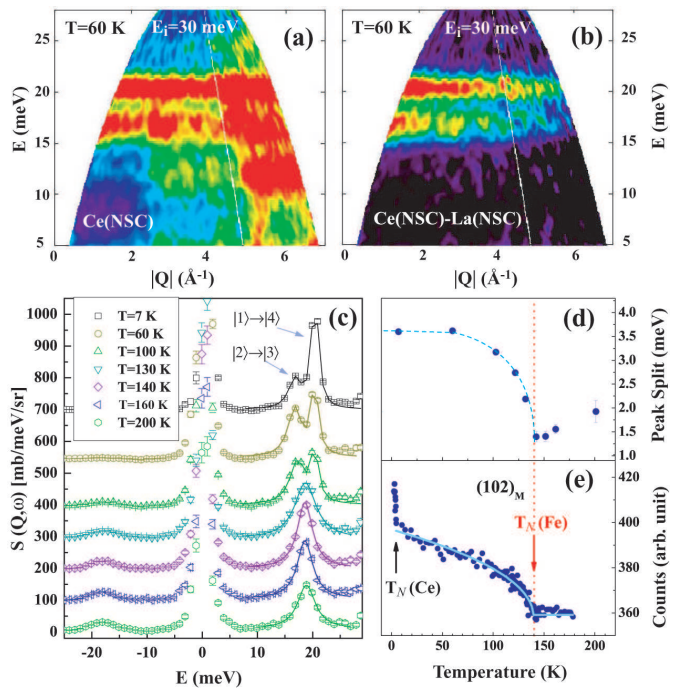}
\caption{(Color online) (a) Raw $S(Q,\omega)$ spectra of CeFeAsO at 60 K and $E_i=30$ meV on MERLIN. (b) Ce CEF excitations after subtration of
the LaFeAsO background. (c) Temperature dependence of the 18.7 meV excitations. 
The peaks around 18 meV at 7 K are fitted with 2 lorentzians, the widths of which were fixed for fittings at all higher temperatures.
 Although the peak separation remain finite above $T_N$, the net separation caused by molecular field can be monitored by the change of this separation as shown in (d). A drastic increase of peak separation happens below 140 K, the $T_N$ of Fe as shown in (e).
 }
\end{figure}

Comparing to the superconducting CeFeAsO$_{0.84}$F$_{0.16}$, the Ce CEF excitations in CeFeAsO near 19 and 65 meV 
have clear double peaks at low temperature that become a single peak at 200 K [Figs. 2(a) and (b)]. In addition, the low-energy spectra 
in the inset of Fig. 2(a) shows a clear peak around 0.7 meV that is not present at 171 K.
 To understand this phenomenon, we carried out careful temperature dependence measurements of the $\sim$19 meV CEF excitation.  
Figure 3(a) shows the raw $S(Q,\omega)$ spectra of CeFeAsO collected on MERLIN at 60 K using $E_i=30$ meV. After subracting the phonon scattering background collected using LaFeAsO, the Ce CEF level shows two clear bands of excitations at 16.6 and 20.4 meV [Fig. 3(b)]. Figure 3(c) shows the detailed temperature dependence of the $\sim$19 meV exitations and Figure 3(d) plots splitting of the two low temperature peaks.  Comparison of these figures with the N$\rm \acute{e}$el ordering temperature of CeFeAsO in Figure 3(e) makes it immediately clear that the CEF splitting is due to the long range AF Fe ordering \cite{zhao}. 

\begin{figure}[tbp]
\includegraphics[scale=.4]{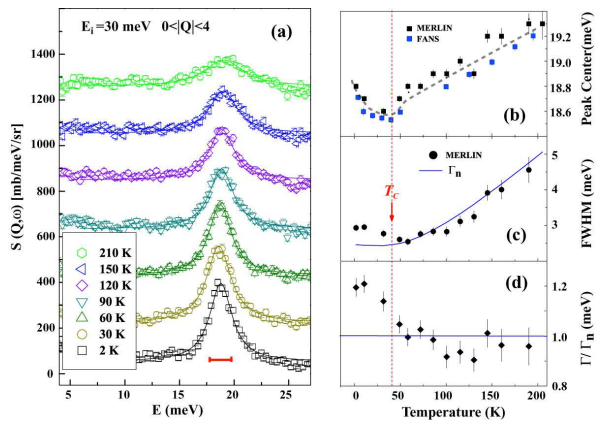}
\caption{(Color online) 
Temperature dependence of the $\hbar\omega_1=18.7$ meV Ce CEF excitation for CeFeAsO$_{0.84}$F$_{0.16}$. (a) MERLIN measurements using $E_i=30$ meV at different temperatures. The instrumental energy 
resolution is 2.1 meV at elastic position (horizontal bar). 
(b) The peak position as a function of temperature for the 18.7 meV CEF level. The solid squres are MERLIN data
while open squares are FANS data, both show clear anomaly at $T_c$. (c) The intrinsic linewidth $\Gamma(T)$ as a 
function of temperature. The solid line shows the expected $\Gamma_n(T)$ assuming noninateracting Fermi liquid.
$\Gamma(T)$ deviates from $\Gamma_n(T)$ near $T_c$.
(d) $\Gamma(T)/\Gamma_n(T)$ shows a clear anomaly near $T_c$, consistent with (c).
 }
\end{figure}

In principle, the orthorhombic structural distortion that precedes the AF ordering in CeFeAsO can have an effect on the Ce CEF levels.
However, since neutron powder diffraction data \cite{zhao} showed that the Ce local environment is not much affected by the lattice distortion, we started fitting the low temperature spectra using the CEF parameters for tetragonal geometry, then added the effect of the molecular field of the Fe spins in the presence of the orthorhombic structural distortion. The solid line in Fig. 2(a) shows our fit to the data, and the CEF parameters for temperatures above and below $T_N$ are also given in Table I.  Assuming that the $\rm \left|I\right\rangle$, $\rm \left|II\right\rangle$, and $\rm \left|III\right\rangle$ doublets in the paramagnetic state of CeFeAsO split into $\left|1\right\rangle$ to $\left|6\right\rangle$ singlets in the AF state as shown in Fig. 1(b), the observed excitations near 19 meV should be composed of 4 possible transitions $\left|1\right\rangle\rightarrow\left|3\right\rangle$,
$\left|1\right\rangle\rightarrow\left|4\right\rangle$, $\left|2\right\rangle\rightarrow\left|3\right\rangle$, and $\left|2\right\rangle\rightarrow\left|4\right\rangle$.  Since the transition probabilities 
$\left|\left\langle 1\right|J_z\left|3\right\rangle\right|^2$ and 
$\left|\left\langle 2\right|J_z\left|4\right\rangle\right|^2$ are rather small, the observed excitations at 
16.8 and 20.4 meV in Fig. 2(a) actually arise from $\left|1\right\rangle\rightarrow\left|4\right\rangle$ and $\left|2\right\rangle\rightarrow\left|3\right\rangle$, and are controlled by the thermal population of $\left|1\right\rangle$ and
$\left|2\right\rangle$, respectively.
This is consistent with the temperature dependence of these two excitations, where the 16.8 mode decreases and the 20.4 meV excitation increases with
decreasing temperature [Fig. 3(c)]. Therefore, the CEF energy levels for $\left|3\right\rangle$ and $\left|4\right\rangle$ should be at $16.6+0.7=17.5$ meV and 20.4 meV, respectively.

\begin{table}[tp]
\caption{Wave functions of different CEF levels for CeFeAsO$_{0.84}$F$_{0.16}$ and CeFeAsO above $T_N$.}
\label{Table}%
\begin{ruledtabular}
\begin{tabular}{cccccccccc}
                             & CeFeAsO$_{0.84}$F$_{0.16}$      & CeFeAsO ($>T_N$)  \\[3pt]
\hline&  \\[3pt]
$\rm \left|I\right\rangle$   & $\left|\mp{1/2}\right\rangle$   &   $\left|\mp{1/2}\right\rangle$    \\[3pt]
$\rm \left|II\right\rangle$  & $-0.581\left|\mp{5\over 2}\right\rangle + 0.814\left|\pm{3\over2}\right\rangle$    &   $-0.445\left|\mp{5\over2}\right\rangle + 0.895\left|\pm{3\over2}\right\rangle$ \\[3pt]
$\rm \left|III\right\rangle$ & $0.814\left|\mp{5\over 2}\right\rangle + 0.581\left|\pm{3\over 2}\right\rangle$     &   
$0.895\left|\mp{5\over2}\right\rangle + 0.446\left|\pm{3\over2}\right\rangle$  \\[3pt]

\end{tabular}
\end{ruledtabular}
\end{table}

In rare-earth substituted high-$T_c$ copper oxides such as Tm$_{0.1}$Y$_{0.9}$Ba$_2$Cu$_3$O$_{6+x}$ \cite{osborn} and   Ho$_{0.1}$Y$_{0.9}$Ba$_2$Cu$_3$O$_{7}$ \cite{boothroyd}, the intrinsic linewidth $\Gamma$ and peak position of the rare-earth CEF excitations are used to probe the local magnetic response of the CuO$_2$ planes and the formation of a superconducting energy gap.
The CEF excitations of Ho and Er have also been used to study pseudogap and order-parameter symmetry in the underdoped superconducting HoBa$_2$Cu$_4$O$_{8}$ and Er$_2$Ba$_4$Cu$_7$O$_{14.92}$, respectively \cite{mesot1}.
To see if the $\hbar\omega_1=18.7$ meV CEF excitation in CeFeAsO$_{0.84}$F$_{0.16}$ is also sensitive to the occurrence of superconductivity, we carefully probed the temperature dependence of its intrinsic linewidth $\Gamma$ [Figs. 4(a), 4(c)] and peak positions [Fig. 4(b)].  Figure 4(a) shows the phonon subtracted data at different temperatures obtained on MERLIN with $E_i=30$ meV. Figure 4(b) plots the peak position as a function of temperature, 
which shows a clear anomaly around $T_c$. 
Inspection of Fig. 4(c) reveals that the linewidth decreases with decreasing temperature.  For the case when the CEF levels do not overlap, the temperature dependence of the linewidth $\Gamma(T)$ of the transition between the ground state ($\left|1\right\rangle$) and the first excited state ($\left|2\right\rangle$) in Fig. 1(c) can be accurately estimated using Eq.(1) in \cite{boothroyd}.  To first order, this can be written as $\Gamma(T)\propto \chi^{\prime\prime}(\hbar\omega_1) \coth(\frac{\hbar\omega_1}{2k_BT})$, where $\chi^{\prime\prime}(\hbar\omega_1)$ is the local integrated magnetic susceptibility at $\hbar\omega_1$ ($=18.7$ meV) and $k_B$ is the Boltzmann constant \cite{osborn}.  To understand $\Gamma(T)$, we calculate the linewidth $\Gamma_n(T)$ below 150 K using this equation 
assuming a noninteracting Fermi liquid normal state \cite{boothroyd}, even though this is still under discussion.
The solid line in Fig. 4(c) shows outcome of our calculation. The observed linewidth starts to deviate from the expected values near $T_c$, as shown more clearly in  
$\Gamma(T)/\Gamma_n(T)$ [Figure 4(d)].
The most natural interpretation of Figs. 3(c) and (d) is that the local integrated magnetic susceptibility 
$\chi^{\prime\prime}(\hbar\omega_1)$ from the Fe spin fluctuations at 18.7 meV increases dramatically below $T_c$.  Since 
the temperature dependence of this excitation and its energy are rather similar to   
that of the neutron spin resonance recently observed in Ba$_{0.6}$K$_{0.4}$Fe$_2$As$_2$ \cite{christianson}, 
we speculate that the observed linewidth change might be related to the spin resonance in CeFeAsO$_{0.84}$F$_{0.16}$.

In summary, we have determined the Ce CEF levles in nonsuperconducting CeFeAsO and superconducting 
CeFeAsO$_{0.84}$F$_{0.16}$ as a function of temperature.  Our results show that Ce CEF excitations are very sensitive to the Fe magnetic order in the undoped system, and dynamic spin susceptibility in superconductor, and therefore can be used as a probe to study spin dynamical properties of 
the rare-earth based oxypnictides.


We acknowledge discussion with B. D. Rainford and thank E. Goremychkin and 
R. Osborn for providing CEF program.
This work is supported by the US DOE BES through DOE DE-FG02-05ER46202, and in part by the US DOE, Division of Scientific
User Facilities, BES. DCS is supported by NSF DMR-0454672. 
The work at the IOP, CAS, is supported by the NSF of China, the CAS 
and the Ministry of
Science and Technology of China.


\end{document}